\documentclass[twocolumn,aps,floatfix,nopacs,superscriptaddress]{revtex4}
\usepackage{amsmath,amssymb,eucal,graphicx}

\DeclareMathOperator{\cotanh}{cotanh}
\DeclareMathOperator{\cosech}{cosech}

\begin{document}

\title{Molecular Spiders with Memory}

\author{Tibor Antal}
\affiliation{Program for Evolutionary Dynamics, Harvard University, Cambridge, MA 02138, USA}
\author{P. L. Krapivsky}
\affiliation{Department of Physics and Center for Molecular Cybernetics, Boston University, Boston, MA 02215, USA}

\begin{abstract}
Synthetic bio-molecular spiders with  ``legs'' made of  single-stranded segments of DNA can move on a surface which is also covered by single-stranded segments of DNA complementary to the leg DNA. In experimental realizations, when a leg detaches from a  segment of the surface for the first time it alters that segment, and legs subsequently bound to these altered segments more weakly. Inspired by these experiments we investigate spiders moving along a one-dimensional substrate, whose legs leave newly visited sites at a slower rate than revisited sites. For a random walk (one-leg spider) the slowdown does not effect the long time behavior. For a bipedal spider, however, the slowdown generates an effective bias towards unvisited sites, and the spider behaves similarly to the excited walk. Surprisingly, the slowing down of the spider at new sites increases the diffusion coefficient and accelerates the growth of the number of visited sites.  
\end{abstract}

\maketitle

\section{Introduction}

Chemists have recently constructed various synthetic molecular systems (see e.g. 
\cite{seeman,pierce,DNA_motor} and a review \cite{review}) which can move on surfaces and tracks. One class of such objects aptly named spiders \cite{exp} consists of relatively small molecules with legs which are short single-stranded segments of DNA. These spiders can move on a surface covered with single-stranded DNA segments, called substrates. The substrate DNA is complementary to the leg DNA. The motion proceeds as legs bind to the surface DNA through the Watson-Crick mechanism, then dissociate, then rebind again, etc.  More precisely, a bond on the substrate with an attached leg is first cleaved \cite{exp}, and the leg then dissociates from the affected substrate (which is called product \cite{exp}). After that the leg can rebind to the new substrate or to the product.

The rate of attachment of a leg of a spider to the substrate and the rate of detachment from the substrate are different from the corresponding rates involving the product. Therefore for the proper description of the motion of a single spider one must keep track of its entire trajectory. This memory requirement makes the problem non-Markovian \cite{markov,van}. Such problems tend to be intractable, though in one dimension non-Markovian problems are occasionally solvable, see e.g. \cite{perman,zerner,ourcookie,volkov,bridge}. 

In this paper we continue the investigation of molecular spiders that move along one-dimensional tracks \cite{mainspider}. The goal is to study the interplay between the spiders and their environment which has been ignored in Ref.~\cite{mainspider}. We shall assume that each leg of a spider steps independently and symmetrically to neighboring empty sites at rates 1 from revisited sites and at rates $r$ from sites visited for the first time. For molecular spiders $r<1$ since legs are bound more firmly to new sites. Two types of spiders were introduced in \cite{mainspider} which differ in the type of constraint that keeps the legs close to each other. For {\bf local spiders} or centipedes, the distance between {\em adjacent} legs is $\leq s$; for {\bf global spiders}, the distance between {\em any} two legs is $\leq S$. Without memory ($r=1$), spiders with local constraint can be mapped onto exclusion processes with open boundaries, while spiders with global constraint map to the simple exclusion process with periodic boundary conditions \cite{mainspider}. These mappings simplify the computations of the diffusion coefficient and, e.g., they allow to determine the diffusion coefficient for general global spiders with an arbitrary number of legs $L$ and an arbitrary maximal distance $S$, and for local $L$-leg spiders with $s=2$.

The non-Markovian character does not make the problem hopeless since the visited area is very simple, namely the visited sites (product) form an island in the sea of unvisited sites. (This happens only in one dimension, and the two-dimensional problem appears intractable.) Graphically, we display a leg of the spider by $\bullet$, an empty sites as $\circ$, and we put a hat at sites which have never been visited earlier, i.e.\ where the DNA substrate is still un-cleaved. A typical configuration for a bipedal spider then looks like
\begin{equation*}
%\label{conf2legs}
 \ldots \hat\circ\,  \hat\circ\,   \hat\circ\,   \hat\circ \underbrace{\circ \circ \circ \circ \circ \circ \circ \bullet \circ }_{\rm island} \hat\bullet\,  \hat\circ\,  \hat\circ\,  \hat\circ\,  \hat\circ \ldots
\end{equation*}

The memory solely acts through the slowdown of legs on newly visited sites. For a one-leg spider, which is merely a random walk, memory does not affect the asymptotic behavior of the number of visited sites and other quantities. On the other hand, the slowdown generates an effective bias of multi-pedal spiders. This bias always points outward. Remarkably,  the slowdown effectively speeds the spider up, that is, it leads to the increase of the diffusion coefficient and the number of visited sites. 

To mimic the effective bias of the spiders, we investigate a random walk which jumps asymmetrically from newly visited sites. This model is similar to the so-called exited walk which has been actively investigated in recent years in the mathematics literature, see e.g. \cite{perman,zerner,ourcookie} and references therein. 

The outline of this paper is as follows. In Sec.~\ref{1leg}, we analyze a random walk 
with memory. First we analyze the symmetric random walk that slows down on the new sites, and then we obtain exact results  for the exited walk which additionally takes into account a bias on the new sites. In Sec.~\ref{2legs}, we discuss the behavior of the bipedal spiders with maximal leg distance $s=2$. Some of the properties of these spiders are exactly computed, while others are established via the analogy with the excited random walk. In Sec.~\ref{multi}, we generalize these results to global spiders with $L=S$. We draw conclusions in Sec.~\ref{conc}. Finally, in an appendix, some of the details required for calculations of Sec.~\ref{1leg} are presented.

\section{Excited Random Walk}
\label{1leg}

In this section we consider the case of a one-leg spider. We shall begin with a nearest-neighbor symmetric random walk on the one-dimensional lattice. By definition, if the walker occupies a site for the first time, the hopping rate to each of its nearest neighbors is $r$; if the walker has already visited this site in the past, it hops at rates 1 to the left and to the right. 

\subsection{Average time to cover $N$ sites}
\label{TN}

A quantity which is most readily computable is the average time $\langle T_N\rangle$ for the random walk to cover $N$ sites. The total number of distinct sites visited by the random walk grows with time, and let $T_N$ be the time when the number of distinct sites first reaches $N+1$ (here $N$ is the size of the island and the $1$ accounts for the site with the random walker on it). Graphically,
\begin{equation*}
\ldots \hat\circ\,  \hat\circ\,   \hat\circ\,   \hat\circ \underbrace{\circ \circ \circ \circ \circ \circ \circ\, \circ}_{N} \hat\bullet\,  \hat\circ\,  \hat\circ\,  \hat\circ\,  \hat\circ \ldots
\end{equation*}
The walk starts at some site and hence $T_0=0$; all the following times $0<T_1<T_2<\ldots$ are random quantities. We shall actually compute the first two moments of
$T_N$. The first moment (the average) is 
\begin{equation}
\label{TN-av}
\langle T_N\rangle=\frac{N(N-1)}{4}+\frac{N}{2r}
\end{equation}

To derive \eqref{TN-av} we first recall how to compute the mean exit
time  from  the interval $(0,M)$  for the random walker hopping at unit rates to  the right and to the left. By definition, the exit time $T(x)$ is the time for the random walker to reach any of the two boundaries, $0$ or $M$, given that  the walker starts at site $x$. The procedure (explained e.g. in \cite{fpp}) is to analyze changes in an infinitesimal time interval $dt$. One writes
\begin{equation*}
%\label{Tx-exact}
T(x) = dt+\left\{
\begin{array}{cl} \displaystyle
T(x)     & \mbox{~~prob~~} 1-2dt\\  \displaystyle
T(x+1) &  \mbox{~~prob~~} dt\\
T(x-1) &  \mbox{~~prob~~}  dt
\end{array}
\right.
\end{equation*}
and making the averaging one arrives at
\begin{eqnarray}
\label{Tx}
\langle T(x)\rangle&=&dt + (1-2 dt)\langle T(x)\rangle\nonumber\\
&+&dt[\langle T(x+1)\rangle+\langle T(x-1)\rangle] 
\end{eqnarray}
for the  mean exit time, which simplifies to
\begin{equation}
\langle T(x+1)\rangle - 2\langle T(x)\rangle + \langle T(x-1)\rangle = -1
\end{equation}
Solving this equation subject to the boundary conditions $T(0)=T(M)=0$ we obtain
\begin{equation}
\label{Tx-sol}
\langle T(x)\rangle =\frac{x(M-x)}{2}
\end{equation}

Returning to our problem, let us compute the time interval $\tau_N$
during which the number of visited sites jumps from $N+1$ to
$N+2$. Without loss of generality we can assume that at $T_N$ the
visited sites are $1,\ldots,N+1$, and the random walk is at site
$N+1$. Similarly to \eqref{Tx} we write an equation for the mean
duration
\begin{equation}
\label{tau-N}
\langle \tau_N\rangle =(1-2rdt)\langle \tau_N\rangle +rdt\,N+dt
\end{equation}
The factor $N$ in the middle term on the right-hand side was derived
by taking into account that if the random walker hops to the left to site $N$, 
the number of distinct sites will increase when the random walk
reaches either site $0$ or $N+2$. This latter problem is equivalent to the above exit problem after identifying $M=N+2$ and $x=N$, and \eqref{Tx-sol} shows that the mean
exit time is $N$. {}From Eq.~\eqref{tau-N} we obtain
\begin{equation}
\label{tau-N-sol}
\langle \tau_N\rangle = \frac{1}{2r}+\frac{N}{2}
\end{equation}
Using
\begin{equation}
\label{Tt}
\langle T_N\rangle=\sum_{j=0}^{N-1}\langle \tau_j\rangle
\end{equation}
in conjunction with \eqref{tau-N-sol} we arrive at \eqref{TN-av}.

If the hopping rates were equal to unity everywhere (no difference
between the substrate and the product), we would have 
\begin{equation}
\label{TN-average}
\langle T_N\rangle_{\rm uniform}=\frac{N(N+1)}{4}
\end{equation}
instead of \eqref{TN-av}. Comparing these two results we see that only
the sub-leading corrections  (linear in $N$)  are different. Hence the
non-Markovian nature has an asymptotically negligible influence on
$\langle T_N\rangle$ for the random walk. 

Other calculations have led to the same conclusion. For instance, we have computed  
$\langle T_N^2\rangle$ and we found 
\begin{eqnarray}
\label{TN2-av-asymp}
\langle T_N^2\rangle  &=& \frac{1}{12}\,N^4+\left(\frac{1}{4r}-\frac{1}{12}\right)N^3\nonumber\\
&+&\left(\frac{1}{6}-\frac{1}{4r}+\frac{1}{4 r^2}\right)N^2
+\left(\frac{1}{4r^2}-\frac{1}{6}\right)N
\end{eqnarray}
For $r=1$ (no memory) equation \eqref{TN2-av-asymp} becomes
\begin{equation}
\label{TN2-av}
\langle T_N^2\rangle_{\rm uniform} = \frac{1}{12}\,N^4+\frac{N^3+N^2}{6}+\frac{1}{12}\,N
\end{equation}
i.e. it has the same leading behavior.

In the $r\ll 1$ limit, memory is very important in the earlier stage;
the crossover time $t_c$ after which memory becomes irrelevant is
estimated by equating the linear and quadratic (in $N$) terms in
\eqref{TN-av}.  This gives
\begin{equation}
\label{time-cross}
t_c\approx r^{-1}
\end{equation}
Thus the memory does not affect the asymptotic behavior of $\langle
T_N\rangle$  for the nearest-neighbor random walk. 

This conclusion may be affected by the gait of the walk. Unfortunately, even for the random walk it is hard to probe complicated gates when we take memory into account. For instance, if the random walk makes $\pm 1, \pm 2$ hops, say all
at the same rates, the problem seems intractable. The above
approach allows to compute $\langle T_N\rangle$ only when the random
walk hops to nearest-neighbor sites or gets back to
its original site. For such gait we found
\begin{equation}
\label{TN-av-mod}
\langle T_N\rangle=\frac{N^2}{4}+\left(\frac{1}{3r} -
\frac{1}{12}\right)N
\end{equation}
hence the influence of
memory on  $\langle T_N\rangle$ for this random walk with modified
gait is again asymptotically irrelevant.

\subsection{Growth of the average number of visited sites}
\label{1legarea}

We now turn to a more general excited walk; we shall use these results for the bipedal spider in Sec.~\ref{2legarea}. An excited walk is defined as follows: When the walker leaves a particular site for the first time it jumps at rate $f$ forward (to unvisited sites) and at rate $b$ backward (to revisit sites). When the walk jumps from a site it had already visited earlier, it jumps at rates one in both directions. For concreteness we define its first ever jump from the origin to the right at rate $f$ and to the left at rate $b$. This special rule, however, is irrelevant in the long time limit which is our main focus. 

The above definition slightly differs from the definition of the excited random walk studied in Refs.~\cite{perman,zerner,ourcookie} where it is a discrete time process, and where the excited walker is biased always into one specific direction (say always to the right).

On the already visited sites, the walker performs a simple symmetric random walk. In the following analysis, we will need certain results for the simple random walk. Particularly, for the walker on the interval $[0,k]$ we will need the exit probability density $g_k(t|m)$ that the walker which starts at site $m$ inside the interval, $0<m<k$, exits at any of the two end sites of the interval during the time interval $(t,t+dt)$. An elementary derivation of this probability density (more precisely, of its Laplace transform) is given in Appendix \ref{exit}.

One can try to repeat the calculation of $\langle T_N\rangle$ in this more general situation. Instead, we consider a dual quantity, namely we shall compute how the average number of visited sites increases with time. Suppose that exited walk had already visited $k-2$ sites and it has just jumped to a new site. Let the already visited sites be $1,\dots,k-2$, and let the random walker be currently at $x=k-1$. (The latter assumption is acceptable since the rules are symmetric. This is not true for the excited random walk of Refs.~\cite{perman,zerner,ourcookie}.) The probability density that the walker will visit the next new site (either site 0 or site $k$) exactly $t$ times later is
\begin{equation}
\label{deltat}
 Q_k(t) = f e^{-(b+f)t} + b \int\limits_0^\infty e^{-(b+f)t'} g_k(t-t'| 2)\, dt'
\end{equation}
The first term in this expression accounts for a direct jump forward from site $k-1$ to $k$, since the walk stays at its original position $k-1$ for time $t$ with probability $\exp[-(b+f)t]$. The second term describes the walk stepping backwards to site $k-2$ after time $t'$, doing a symmetric walk on sites $1,\dots,k-1$, and exiting this interval at either ends at time $t$ with probability density $g_k(t-t'| 2)$. The Laplace transform of \eqref{deltat} is
\begin{equation}
\label{deltatlap}
 Q_k(s) = \frac{f+bg_k(s|2)}{s+b+f}
\end{equation}
[The Laplace transform of the general exit probability density $g_k(s| m)$ is given by \eqref{exitlr}.]

The above argument works also for the first ever step of the walk with our definition. If we prefer the assumption of hopping at rate $f$ in both directions for the first time, we have to use $Q_2(s)=2f/(s+2f)$. This, however, does not affect the long time behavior.

The probability density that the walk visits the $n^{\rm th}$ new site at time $t$ is 
\begin{equation}
\label{conv}
 F_n(t) = \int \prod_{k=2}^n Q_k(t_k)\, dt_k
\end{equation}
where $t_2+t_3+\dots+t_n=t$. The Laplace transform of the convolution \eqref{conv} is
\begin{equation}
\label{Fdeflap}
 F_n(s) = \prod_{k=2}^n Q_k(s)
\end{equation}

One anticipates that the main contribution in the long time limit comes from terms with $k\sim\sqrt{t}$. This suggests to take the $s\to 0$ limit while keeping $\sqrt{s}k$ finite. In this limit $g_k(s|m)$ is given by \eqref{symlim} and hence \eqref{deltatlap} becomes
\begin{equation}
\label{Qsmalls}
 Q_k(s) = 1 - a\sqrt{s}\, \tanh\frac{\sqrt{s}k}{2}
\end{equation}
where we have used the shorthand notation
\begin{equation}
\label{afb}
 a = \frac{2}{1+f/b}
\end{equation}
Note that $a$ and consequently the whole distribution $F_n(s)$ only depends on the ratio $f/b$ of the hopping rates. The no bias case corresponds to $a=1$, which coincides with any symmetric hopping $f=g$. This has an interesting consequence --- if the walk is only slowed down at new sites (but not biased), then it has no effect on its long time behavior, in agreement with the conclusions of the previous subsection \ref{TN}. Note also that $0\le a\le2$.

Now we return to the calculation of $F_n(s)$. The product in \eqref{Fdeflap} can be re-written as
\begin{equation}
\label{Fsumlap}
 F_n(s) = \exp \left( \sum_{k=2}^n \ln Q_k(s) \right) 
\end{equation}
In the $s\to 0$ limit we use \eqref{Qsmalls} to obtain
\begin{equation}
 F_n(s) = \exp \left( - a\sqrt{s} \sum_{k=2}^n   \tanh\frac{\sqrt{s}k}{2} \right) 
\end{equation}
In the same limit the sum becomes an integral
\begin{equation}
 \int\limits_0^n  \tanh\frac{\sqrt{s}k}{2} \, dk 
 =  \frac{2}{\sqrt{s}}  \ln \cosh \frac{\sqrt{s}n}{2}  
\end{equation}
and hence
\begin{equation}
 F_n(s) = \left( \cosh \frac{\sqrt{s}n}{2} \right)^{-2a}
\end{equation}

At this point we recognize that this generating function is identical (up to a change in notation) to the one obtained in \cite{ourcookie} for a discrete time excited walk on the half infinite line. This allows us to use the results of Ref.~\cite{ourcookie}. 
%Hence we shall use the results obtained there by replacing their $q$ to $a$, and their $n$ to $n\sqrt{2}/4$.

Note that $F_n(s= 0)=\int_0^\infty F_n(t)dt=1$ since the walk will cover any number of sites with probability one. The cumulants \cite{van} can be directly obtained from the generating function $F_n(s)$ to yield
\begin{equation}
 \kappa_\ell = a\, C_\ell\, n^{2\ell}
\end{equation}
with
\begin{equation}
 C_\ell = 8^{1/3} (1-2^{-2\ell})(\ell-1)! \pi^{-2\ell}\zeta(2\ell)
\end{equation}
For example, $C_1=1/4$, $C_2=1/48$, $C_3=1/240$. The average time to cover $n$ sites, and its fluctuation, are
\begin{equation}
\label{cums}
 \kappa_1=\langle t \rangle =  \frac{a n^2}{4} ~,~~~ \kappa_2=\langle t^2 \rangle -  \langle t \rangle^2 = \frac{a n^4}{48}
\end{equation}
Note that these exact expressions have been already derived and given by equations \eqref{TN-av} and \eqref{TN2-av-asymp} in the symmetric $a=1$ case.

%%%%%%%%%%%%
\begin{figure}
\centering
\includegraphics[scale=0.7]{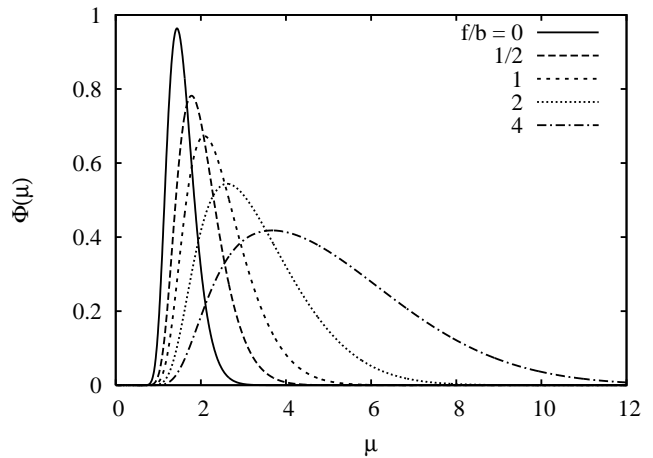}
\caption{The scaled distribution  \eqref{Phiex} of the number of visited sites by the exited walk for several values of bias.}
\label{exL1dis}
\end{figure}
%%%%%%%%%%%%%

The distribution $F_n(t)$ attains a scaling form
\begin{equation}
 F_n(t) = t^{-1}\Phi(\mu)\,,\quad \mu=\frac{n}{\sqrt{t}}
\end{equation}
in the scaling limit $n,t\to\infty$ with the scaling variable $\mu$ kept finite. The scaled 
distribution $\Phi(\mu)$ has been computed in Ref.~\cite{ourcookie}, which in the present notation reads
\begin{equation}
 \Phi(\mu) = \mu \frac{2^{2a-1}}{\sqrt{\pi}} \sum_{k=0}^\infty 
 \binom{-2a}{k} (k+a) \, e^{-\mu^2 (k+a)^2/4} 
 \label{Phiex}
\end{equation}

The average number of visited sites at time $t$ is 
\begin{equation}
 \langle n \rangle = \frac{\int_0^\infty n F_n(t)\, dn}{\int_0^\infty F_n(t)\, dn} = A(a)\sqrt{t}
\end{equation}
with
\begin{equation}
\label{Phinorm}
A(a) = 2\int\limits_0^\infty d\mu~ \Phi(\mu) = \frac{2 \Gamma(a)}{\Gamma(a+1/2)}
\end{equation}
where we have integrated \eqref{Phiex} term by term, and used the duplication formula for Gamma functions. Note that $A(a)$ is a monotonically decreasing function for $a>0$. (This can be proved e.g. by differentiating $\ln A(a)$ and using the properties of the di-gamma function.)

In the no-bias case ($a=1$), equation \eqref{Phinorm} reduces to the well known result \cite{weiss} for the amplitude $A(1)=4/\sqrt{\pi}$. In the limits of strong forward and backward biases, the amplitude behaves as
\begin{equation}
A(a) = \left\{
\begin{tabular}{ll}
$\displaystyle \frac{2}{a\sqrt\pi} + O(1)$           & for $a\ll 1$\\ \\
$\displaystyle \frac{8}{3\sqrt\pi} + O(2-a)$ ~~~ &  for $2-a\ll1$
\end{tabular}
\right.
\end{equation}
Note that the naive inversion of the average time \eqref{cums} would result in the incorrect $A=2/\sqrt{a}$.

%%%%%%%%%%%%%%%%%%%%%%%%%%%
\section{Bipedal Spider}
\label{2legs}

Here we compute various characteristics of the simplest bipedal spider with $s=2$. We shall see that memory affects the leading behaviors, and we will explain these findings 
by showing that the slowdown of legs on new sites leads to effective outward bias when the spider is at the boundary of the island of visited sites. 

\subsection{Average time to cover $N$ sites}

Since the spider configuration is $\bullet\circ\hat\bullet$ when a new site is first visited, it is convenient to start from the same type of configuration 
\begin{equation*}
\ldots \hat\circ\,  \hat\circ\,   \hat\circ\,   \hat\circ \bullet \circ\, \hat\bullet\, \hat\circ\,  \hat\circ\,  \hat\circ\,  \hat\circ \ldots
\end{equation*}
This assumption simplifies the calculation, but does not alter the asymptotic behavior.

Denote by $\tau_N$ the time interval during which the number of visited sites jumps from $N+3$ to $N+4$; then the number of visited sites first reaches $N+3$ at the moment $T_N=\sum_{0\leq j\leq N-1}\tau_j$. We can assume that at $T_N$ the visited sites are
$1,\ldots,N+3$, the spider's legs are at sites $N+1$ and $N+3$. Thus the configuration at time $T_N$ is
\begin{equation}
\label{config}
\ldots\hat\circ\, \hat\circ\, \hat\circ \underbrace{\circ\cdots\circ}_{N}\bullet\circ\hat\bullet\, \hat\circ\, \hat\circ\, \hat\circ \ldots
\end{equation}
Similarly to \eqref{tau-N} we write
\begin{eqnarray}
\label{tau-N-spider}
\langle \tau_N\rangle &=&  [1-(1+r)dt]\langle \tau_N\rangle
+rdt\,\frac{3(2N+1)}{2}\nonumber\\ 
&+&dt\,\langle \widehat{\tau}_N\rangle+dt
\end{eqnarray}
for the mean duration. The second term on the right-hand side accounts
for the right leg of the spider hopping from  site $N+3$ to $N+2$. The
site $N+3$ turns from the substrate into the product and the
configuration becomes
\begin{equation*}
\ldots\hat\circ\, \hat\circ\, \hat\circ\, \underbrace{\circ\cdots\circ}_{N}\bullet\bullet\circ\, \hat\circ\, \hat\circ\, \hat\circ \ldots
\end{equation*}
We term the mean position of the legs as the "center of mass" of the spider.
Note that the center of mass of the spider is at position $N+3/2$ and it makes $\pm 1/2$ hops with unit rates. The visit of sites $1$ or $N+3$ by the center of mass corresponds to visiting sites $0$ or $N+4$ by the spider. The mapping to the random walk allows us to use \eqref{Tx-sol} with $M=2N+4$ and $x=2N+1$ to yield $3(2N+1)/2$ for the average time.

The third term on the right-hand side of Eq.~\eqref{tau-N-spider}
accounts for the left leg of the spider hopping from site $N+1$ to
$N+2$. Starting from this position, the mean duration time $\langle
\widehat{\tau}_N\rangle$ before a new site is visited is found by
writing an equation similar to Eq.~\eqref{tau-N-spider}
\begin{equation}
\label{tau-N-sp}
\langle  \widehat{\tau}_N\rangle =  [1-(1+r)dt]\langle
\widehat{\tau}_N\rangle +  dt\,\langle \tau_N\rangle+dt
\end{equation}
Simplifying Eqs.~\eqref{tau-N-spider}--\eqref{tau-N-sp} we obtain
\begin{subequations}
\begin{align}
&(1+r)\langle \tau_N\rangle = r\,\frac{3(2N+1)}{2} + \langle \widehat{\tau}_N\rangle+1
\label{tt1}\\
&(1+r)\langle \widehat{\tau}_N\rangle = \langle \tau_N\rangle+1
\label{tt2}
\end{align}
\end{subequations}
from which we find
\begin{equation}
\label{tau-N-spider-sol}
\langle \tau_N\rangle = \frac{1}{r}+\frac{3(2N+1)}{2}\,\frac{1+r}{2+r}
\end{equation}
Plugging \eqref{tau-N-spider-sol} into the sum \eqref{Tt} we obtain
\begin{equation}
\label{TN-av-spider}
\langle T_N\rangle=3\,\frac{1+r}{2+r}\,\frac{N^2}{2}+\frac{N}{r}
\end{equation}

If there is no difference between the substrate and the product and the hopping rates are equal to $r$ everywhere, the average time is equal to
\begin{equation}
\label{TN-average-spider}
\langle T_N\rangle_{\rm uniform}=N(N+1)
\end{equation}
Hence memory is important for spiders as it affects the leading asymptotic behavior of $\langle T_N\rangle$.

\subsection{Effective bias}

We now argue that a slowdown at new sites leads to an effective bias for bipedal spiders. Imagine that the spider is in configuration \eqref{config}. Analyzing various routes, we obtain that the right leg first hops to the left with probability
\begin{equation}
\label{q-eq}
q=\frac{r}{1+r}+\frac{1}{1+r}\,\frac{1}{1+r}\,q
\end{equation}
and otherwise it hops to right with probability $p=1-q$. An explanation of \eqref{q-eq} is as follows.  The first term on the right-hand side describes the probability that the first leg which is going to hop will be the right leg --- this happens with probability $\frac{r}{1+r}$ and the leg ought to hop to the left. If the first leg to hop is the left leg --- this happens with probability $\frac{1}{1+r}$ --- it then has to hop back (the probability is again $\frac{1}{1+r}$) and then we are back to the initial configuration thereby providing us with factor $q$. Solving \eqref{q-eq} we find 
\begin{equation}
\label{pq-sol}
q=\frac{1+r}{2+r}\,,\quad p=\frac{1}{2+r}
\end{equation}

The probabilities $p$ and $q$ are somewhat misleading, e.g. if the right leg first hops to the left, the spider configurations are different --- in the starting one $\bullet\circ\hat\bullet\, $ the legs are separated, while in the final configuration $\bullet\bullet\circ$ the legs are in adjacent sites. It is easier to appreciate the probabilities $p_+$ and $p_-$ that the spider first moves forward and backward, respectively, while the configuration remains the same.

Let us rephrase this definition. First note that knowing the position $x$ of the center of mass fully determines the position of both legs of the bipedal spider with $s=2$. We assign $x=0$ to the initial configuration \eqref{config}. The jump of any leg changes the position of the center of mass by $\pm 1/2$, hence $x$ takes half-integer values. We are interested in the probability $p_+$ that the center of mass moves forward by one lattice site to $x=1$ first, and in the probability $p_-=1-p_+$ that it moves backward to $x=-1$ first.

We already know the probabilities of the first jump of the right leg \eqref{pq-sol}. With probability $p$ the right leg jumps to the right first, and the spider is already at $x=1$. On the other hand, the right leg jumps to the left first with probability $q$, which puts the spider at $x=-1/2$. The spider then performs a simple random walk on the half integers,  and reaches $x=1$ first with probability $1/4$. Collecting these terms we arrive at
\begin{equation}
\label{p+}
 p_+ = p + \frac{q}{4} = \frac{1}{2} + \frac{1-r}{4(2+r)}
\end{equation}
and $p_-=1-p_+$. Thus the spider moves symmetrically when $r=1$ (no memory), and as $r$ decreases, the bias {\em forward} increases monotonically. The bias is the strongest in the $r\to 0$ limit when $p_-= 3/8$ and $p_+=5/8$.

\subsection{Growth of the average number of visited sites}
\label{2legarea}

We now apply the above results to the bipedal spider. We approximate our spider by an effective walker that steps from already visited sites at rate one to both directions. The step length of this effective particle is $1/2$. Therefore it visits only half as many sites of the original lattice as a particle whose step length is one. Thus 
$\langle N \rangle = A_2(r)\,\sqrt{t}$ with 
\begin{equation}
\label{A_2}
 A_2(r) = \frac{A(r)}{2}
\end{equation}
Without memory we have $A_2(r=1)=2/\sqrt\pi$.

\begin{figure}
\centering
\includegraphics[scale=0.7]{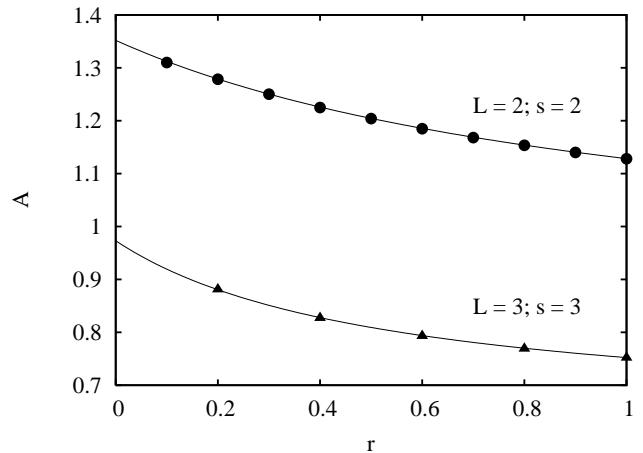}
\caption{The amplitude $A$ in the growth law $ \langle N \rangle = A\sqrt{t}$ as a function of $r$. For the bipedal $L=2$ spider with maximal leg distance $s=2$, simulations are in excellent agreement with the theoretical prediction \eqref{Ar}. The same is also valid for the $L=3$ spider with $S=3$, where the theoretical prediction is given by \eqref{AL}.}
\label{fig:amp}
\end{figure}

{}From a new site, the effective particle hops predominantly in the forward direction.  We approximate the ratio of the forward and backward rates of the effective particle by the ratio of the forward and backward probabilities, $f/b=p_+/p_-$. Using \eqref{p+} we conclude that parameter $a$, see \eqref{afb}, becomes 
\begin{equation}
\label{ar}
 a = \frac{3(1+r)}{2(2+r)}
\end{equation}
Using \eqref{Phinorm} and \eqref{A_2}--\eqref{ar} we arrive at the prediction for the average number of visited sites at time $t$:
\begin{equation}
\label{Ar}
 \langle N \rangle = A_2(r)\,\sqrt{t}\,,\quad
 A_2(r)=\frac{\displaystyle \Gamma\left(\frac{3+3r}{4+2r}\right)}{\displaystyle \Gamma\left(\frac{5+4r}{4+2r}\right)}
\end{equation}
The amplitude $A_2(r)$ depends only weakly on $r$ in the $0<r\le1$ region. As $r$ increases from 0 to 1, $A_2(r)$ decreases monotonically from 
\begin{equation*}
A_2(r=0) = \frac{\Gamma(3/4)}{\Gamma(5/4)} \approx 1.3519
\end{equation*}
to $A_2(r=1)=2/\sqrt\pi\approx 1.1283$.
Fig.~\ref{fig:amp} shows an excellent (perhaps exact) agreement between the above theory and simulation results.

%%%%%%%%%%%%%%
%%%%%%%%%%%%%%

%%%%%%%%%%%%%%
\begin{figure}
\centering
\includegraphics[scale=0.7]{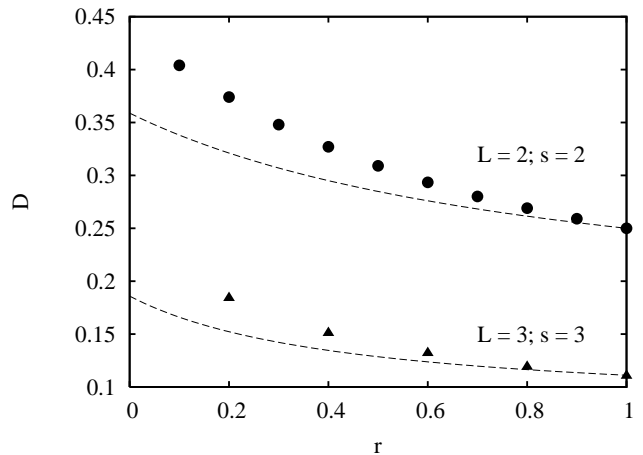}
\caption{Simulation results for the diffusion coefficient as a function of rate $r$ for the bipedal $L=2$ spider with maximal leg distance $S=2$, and for the $L=3$ spider with $S=3$. The dashed lines are the approximation \eqref{Dapp}.}
\label{fig:diff}
\end{figure}
%%%%%%%%%%%%%%

\subsection{Diffusion coefficient}

Up to now we have probed the spider's motion through the trace it leaves --- the number of visited sites. This is the quantity which is measured in experiments \cite{exp}. Interestingly, a direct analysis of the spider motion is much more challenging both experimentally and theoretically. 

One can of course proceed numerically. According to simulations, the position of the spider (averaged over many runs) follows a Gaussian distribution in the large time limit. This is somewhat counterintuitive since the spider slows down at the extreme positions. Overall this is a greatly simplifying feature since if the position is (asymptotically) Gaussian, we can characterize the spider's motion solely by its diffusion coefficient $D$. 

Since the Gaussian distribution gives the position of the spider in the large time limit, one might try to approximate its diffusion coefficient through the relation $\langle N \rangle = 4\sqrt{Dt/\pi}$ which is valid for simple random walks. This leads to 
\begin{equation}
\label{Dapp}
 D \approx \frac{\pi A^2}{16}
\end{equation}
For $r=1$ (no memory), the relation $D=\pi A^2/16$ is exact. For $r<1$,
however, it underestimates the spider's diffusion coefficient, see Fig.~\ref{fig:diff}.

%%%%%%%%%%%%%%%%%%%%%%%%
\section{multi-pedal spiders}
\label{multi}

Spiders with an arbitrary number of legs $L$ are difficult to analyze. Here we consider global spiders which (by definition) satisfy the following constraint --- the distance between any two legs is $\leq S$. Since the site cannot accommodate more than one leg $S\geq L-1$, and the spiders with $S=L-1$ are immobile, so the first interesting case corresponds to $S=L$. These global spiders include the bipedal spiders studied in Sec.~\ref{2legs}. We now show that these global spiders can be analyzed generalizing the approach of Sec.~\ref{2legs}. 

First we notice that knowing the center of mass position of a spider with $S=L$ we can read off the position of all its legs. Hence as for the $L=2$ case we can replace the spider by an effective particle whose position coincides with the center of mass of the spider. The step length of this effective particle is $1/L$. Consider a configuration 
\begin{equation*}
 \ldots \hat\circ\,  \hat\circ\,   \hat\circ\,   \hat\circ \circ \circ \circ \circ \bullet \bullet \bullet \circ\, \hat\bullet\,  \hat\circ\,  \hat\circ\,  \hat\circ\,  \hat\circ \ldots
\end{equation*}
where the right leg has just stepped to a new site. We are interested in the first passage probability $p_+$ for the effective particle to move one whole lattice spacing ($L$ steps) to the right before moving one lattice spacing to the left. Denote again by $q$  the probability that the rightmost leg will first jump to the left. Repeating an argument that led us to \eqref{q-eq}, we arrive at 
\begin{equation}
\label{q-eq-L}
q=\frac{r}{1+r}+\frac{1}{1+r}\,\epsilon\,q
 ~,~~~ \epsilon = \frac{1+(L-2)r}{1+(L-1)r}
\end{equation}
Here $\epsilon$ is the probability that after the right jump of the second leg from the right, the spider returns to the initial configuration without ever moving the rightmost leg. This simple first passage problem can be solved, see e.g. \cite{van}, to obtain \eqref{q-eq-L}. {}Form \eqref{q-eq-L}, the probability that the rightmost leg first jumps to the left is
\begin{equation}
 q = \frac{1+(L-1)r}{2+(L-1)r}
\end{equation}
The probability that the rightmost leg first jumps to the right is $p=1-q$. Following the same reasoning as in the $L=2$ case, we finally arrive at
\begin{equation}
\label{pplus-L}
 p_+ = \frac{1}{2} + \frac{(L-1)(1-r)}{2L[2+r(L-1)]}
\end{equation}
and $p_-=1-p_+$. As $r$ increases from 0 to $\infty$, the bias to the right $p_+$ decreases monotonically, and the motion is unbiased $p_+=1/2$ for the neutral $r=1$ case. Hence, a slowdown at new sites ($r<1$) leads to the outward bias for arbitrary $L$. It is also clear from \eqref{pplus-L} that the bias is weaker for larger spiders.

To calculate the number of visited sites we first obtain
\begin{equation}
\label{a-L}
 a = \frac{L+1}{L}\, \frac{1+(L-1)r}{2+(L-1)r}
\end{equation}
from \eqref{pplus-L}. Recalling \eqref{Phinorm} and taking into account the normalization $A_L=A/L$ we find the amplitude 
\begin{equation}
\label{AL}
  A_L(r)=\frac{2}{L}\,\frac{\Gamma(a)}
 {\Gamma\left(a+1/2\right)}
\end{equation}
in the growth law $\langle N \rangle = A_L(r)\,\sqrt{t}$. An agreement between theory and simulations is excellent, see Fig.~\ref{fig:amp}.

Equation \eqref{a-L} shows that $a=a(r,L)$ is an increasing function of both $r$ and $L$. Since $\Gamma(a)/\Gamma(a+1/2)$ is a decreasing function of $a$, we conclude that for $L=S$ global spiders the number of visited sites decreases as $r$ or $L$ grows. 

We believe that this is generally true for arbitrary spiders, i.e. $A$ decreases as $r$ or $L$ grows. Intuitively it is easy to understand the monotonic $L$ dependence, as an additional leg only blocks the motion of the others. The $r$ dependence of $A$ is more intriguing, since slowing down the spider actually results in more visited sites. One could argue that the slowdown at the boundary keeps the spider around the unvisited sites, hence it visits more of them. This argument, however, is flawed since the slowdown does not effect the behavior of a random walk as we saw it in Sec.~\ref{1leg}. 

Here is an argument that supports the above conjecture. When the rightmost leg steps to a new site it has to stretch out a bit. If legs slow down on new sites, it gives other legs time to catch up, which leads to the outward bias. Another argument is that the when the rightmost leg stays longer at the new site, the second leg has more chance to step next to it, hence forcing the rightmost leg to the right. On the other hand, when the rightmost leg steps to a new site, the spider is a bit stretched out and possibly cannot step further due to some constraint (e.g. local or global). When the rightmost leg stays in place longer, it gives more time to the other legs to step closer, allowing the rightmost leg to step to the right, hence it leads again to a bias to the right.

\section{Conclusion}
\label{conc}

In experimental realizations \cite{exp}, spiders affect the environment which in turn affects their motion. More precisely, the state of any site on a surface changes irreversibly after the first visit of a leg of a spider and then it remains the same. During the subsequent visits legs are bound less firmly to previously visited sites, and we model this feature by prescribing different hopping rates from sites visited in the past and sites visited for the first time. Namely a leg slows down on new sites while the spider remains agile inside a previously visited region. Thus the knowledge of the spider's entire trajectory is required to describe its motion. 

First, we considered a random walk (a one-leg spider) that slows down on the new sites. We calculated first two moments, $\langle T_N\rangle$ and $\langle T_N^2\rangle$, of the time $T_N$ to cover $N$ sites, and also the dual quantity $\langle N \rangle$, that is the average number of sites visited until time $T$. All these quantities exhibit the same {\em leading} asymptotic behavior as in the case of a simple random walk without slowing down. These concrete results clearly reveal the general truth, namely that the non-Markovian nature of the problem affects only sub-leading behaviors of the random walk with slowing down. All exact calculations support this conjecture, and we think that it is valid even in two dimensions. Heuristically, the above hypothesis relies on the crucial feature of a random walk in one and two dimensions --- the random walk keeps visiting the same sites again and again (the average number of visits diverges with time), and therefore the slowing down on new sites becomes asymptotically less and less important. 

The above argument seemingly applies to spiders as well, so it was surprising that the computation of the average time $\langle T_N\rangle$ to cover $N$ sites for the simplest bipedal spider with $s=2$ gave a result that differs from the corresponding result for the same bipedal spider without slowdown. Thus, the slowdown affects the behavior even in the leading order. The especially striking feature is that the spider is effectively more agile, that is, the number of covered sites grows faster than for the same spider without slowdown. The mechanism leading to this acceleration is an effective bias towards unvisited sites. This bias is not apparent since legs hop {\em symmetrically}. 
Yet once the front leg steps onto a new site, it slows down and the hind leg hops to the neighboring site thereby effectively pushing (due to exclusion) the front leg further into the unvisited region. This argument does not fully explain the bias as the hopping processes are stochastic, but it does make the bias plausible. For the bipedal sider with $s=2$, and more generally for the class of multi-pedal spiders with the maximal span equal to the number of legs ($S=L$), we computed the bias analytically. We believe that the bias exists for all spiders; numerically we confirmed this belief in all cases we looked at, but this is still a conjecture which may be very difficult to prove.  

The bias explains the qualitative behavior of the average time $\langle T_N\rangle$, and it suggests that the dual quantity $\langle N\rangle$, that is, the average number of sites visited by time $t$, also grows faster than for spiders without slowdown on the new sites. This quantity is important as it is directly measured in experiments \cite{exp}. We computed $\langle N\rangle$ by replacing the spider by an excited walk, that is a walk that hops symmetrically from previously visited sites but biased towards the unvisited region when hopping from new sites. For this excited walk we calculated not only the average, but the whole probability distribution of the number of visited sites. The replacement of the spider by an excited walk is, however, not  rigorous, but presumably exact due to the perfect agreement with simulations of the spiders.

Apart from $\langle T_N\rangle$ and $\langle N\rangle$, we tried to compute the diffusion coefficient. Numerically we observed that the probability distribution of the spider position approaches a Gaussian distribution in the long time limit, and therefore the diffusion coefficient fully characterizes the probability distribution of the spider position. We neither proved that the probability distribution is (asymptotically) Gaussian nor computed the diffusion coefficient. This is an interesting challenge for future work.
 
Finally we note that models of synthetic molecular spiders are closely related to models of natural molecular motors \cite{motors,det-frey,det-evans,filament,traffic,mauro,cargo}. Memory effects can occur when motors change their environment, which then affects the motors motion. (This happens e.g. for the collagenase which moves along collagen fibrils  \cite{bridge-exp}.) The non-Markovian nature of these situations makes these problems challenging, nevertheless, some problems have been analyzed \cite{bridge,bridge-numer}. It should be possible to employ the methods developed here in the context of molecular spiders, and to imply them to the behavior of molecular motors in situations when memory effects are important.

\acknowledgments{We are very grateful to Kirone Mallick for numerous helpful suggestions. We also acknowledge financial support  to the Program for Evolutionary Dynamics at Harvard University by Jeffrey Epstein (TA), NIH grant R01GM078986 (TA), and NSF grant CHE0532969 (PLK).}

%%%%%%%%%%%%%%%%%%%%%%%%%%%%%%%
\appendix
\section{Escape of a symmetric random walk from an interval}
\label{exit}

We have a random walk on the integers which hops at rate one to the left, and at rate one to the right. The walk starts at time $t=0$ from site $m$, where $1\le m\le k-1$, and site $0$ and $k$ are absorbing. We are interested in the first passage probability distribution to reach either site $0$ or $k$ for the first time.

The position of the walk on the infinite lattice at time $t$ is given by \cite{van}
\begin{equation}
 \label{inf}
 q(n,t|m) = e^{-2t} I_{|n-m|}(2t)
\end{equation}
We then do a mirror trick. We start infinitely many walkers at time $t=0$ from positions $n=2ak+m$, and 'negative' walkers from  $n=2ak-m$, where $a$ runs over all integers. Now their superposition gives the probability distribution of the walker in the interval
\begin{equation}
\label{inter}
 p(n,t|m) = \sum_{a=-\infty}^{\infty}  [q(n,t|2ak+m) - q(n,t|2ak-m)] 
\end{equation}
since, from symmetry reasons, $p(0,t)=p(k,t)=0$ for all time. 

The Laplace transform of the position distribution $q(n,t|m)$ is 
\begin{equation}
 Q(n,s|m) = \frac{\left(s+2-\sqrt{s(s+4)}\right)^{|n-m|}}{2^{|n-m|}\sqrt{s(s+4)}} = \frac{c^{|n-m|}}{d}
\end{equation}
where we have used shorthand notation
\begin{equation}
 c=1+ \frac{s-\sqrt{s(s+4)}}{2} ~,~~~ d=\sqrt{s(s+4)}
\end{equation}
Hence we have to sum up a geometric series to obtain the Laplace transform of \eqref{inter}
\begin{equation*}
 P(n,s|m) = \frac{c^{|n-m|}-c^{n+m}}{d} + \frac{(c^m-c^{-m})(c^n-c^{-n})}{d(1-c^{-2k})} 
\end{equation*}
The first passage probability to site $0$ and site $k$ is simply
\begin{equation}
\begin{split}
\label{exitlr}
 L(s|m) &= P(1,s|m) = \frac{(c-c^{-1})(c^{m-k}-c^{k-m})}{d(c^k-c^{-k})}\\
 R(s|m) &= P(k-1,s|m) = \frac{(c-c^{-1})(c^{-m}-c^{m})}{d(c^k-c^{-k})}
\end{split} 
\end{equation}
respectively, and the probability to leave the interval at either end is $g_k(s|m)=L(s|m)+R(s|m)$.

We are interested in the $s\to 0$, $k\to\infty$ limit, with $k\sqrt{s}$ and $m$ being constants. Noting that in this limit
\begin{equation*}
\begin{split}
 c^m ~&\to~ 1-m\sqrt s + O(s)\\
 c^{-m}-c^{m} ~&\to~ 2m\sqrt s + O(s^{3/2})\\
 d ~&\to~ 2\sqrt s + O(s^{3/2})\\
 c^k ~&\to~ e^{-k\sqrt s} + O(s)\\
\end{split}
\end{equation*}
we establish the expansions 
\begin{equation}
\label{symasym}
\begin{split}
 L(s|m)  ~&\to~ 1-m\sqrt s \cotanh k\sqrt s + O(s)\\
 R(s|m)  ~&\to~ m\sqrt s \cosech k\sqrt s + O(s)
\end{split}
\end{equation}
Therefore the exit at either side of the interval becomes
\begin{equation}
\label{symlim}
g_k(s|m) ~\to~ 1-m\sqrt s \tanh \frac{k\sqrt s}{2} + O(s)
\end{equation}
This asymptotic has been used in subsection \ref{1legarea}.

Finally we notice that the diffusion approximation \cite{fpp} of this simple random walk
\begin{equation*}
 \tilde L(s|m)  = \frac{\sinh (k-m)\sqrt s}{\sinh k\sqrt s} ~,~~
 \tilde R(s|m)  = \frac{\sinh m\sqrt s}{\sinh k\sqrt s}
\end{equation*}
leads to the exact asymptotic behavior \eqref{symasym} in the leading order. This property has been exploited in \cite{ourcookie}.


\begin{thebibliography}{99}
\bibitem{markov} For Markov processes, the future is determined by the
   present \cite{van}.
\end{thebibliography}
\end{document}